 \documentclass[aps,prb,twocolumn,groupedaddress]{revtex4}
\usepackage{graphicx}

\begin{document}

\title{Magnetic properties of $\sigma$-FeCr alloy as calculated with the
charge and spin self-consistent KKR(CPA) method}

\author{J. Cieslak}
\email[Corresponding author: ]{cieslak@novell.ftj.agh.edu.pl}
\author{J. Tobola}
\author{S. M. Dubiel}
\author{W. Sikora}
\affiliation{Faculty of Physics and Applied Computer Science,
AGH University of Science and Technology, al. Mickiewicza 30, 30-059 Krakow, Poland}

\date{\today}

\begin{abstract}
Magnetic properties of a $\sigma-$Fe$_{16}$Cr$_{14}$ alloy calculated with the charge and spin self-
consistent Korringa-Kohn-Rostoker (KKR) and combined with coherent potential approximation (KKR-CPA) methods are reported. Non-magnetic state as well as various magnetic
orderings were considered, i.e. ferromagnetic (FM) and more complex anti-parallel (called
APM) arrangements for selected sublattices, as follows from the symmetry analysis. It has
been shown that the Stoner criterion applied to non-magnetic density of states at the Fermi energy, $E_F$ is
satisfied for Fe atoms situated on all five lattice sites, while it is not fulfilled for all Cr atoms.
In FM and APM states, the values of magnetic moments on Fe atoms occupying various sites
are dispersed between 0 and 2.5 $\mu_B$, and they are proportional to the number of Fe atoms in
the nearest-neighbor shell. Magnetic moments of Cr atoms havin much smaller values were found to be coupled
antiparallel to those of Fe atoms. The average value of the magnetic
moment per atom was found to be $<\mu>=0.55 \mu_B$ that is by a factor of 4 larger than the
experimental value found for a $\sigma-$Fe$_{0.538}$Cr$_{0.462}$ sample. Conversely, admitting an anti-
parallel ordering (APM model) on atoms situated on C and D sites, according to the group theory and
symmetry analysis results, yielded a substantial reduction of $<\mu>$ to 0.20 $\mu_B$. Further
diminution of $<\mu>$ to 0.15 $\mu_B$, which is very close to the experimental value of 0.14 $\mu_B$, has
been achieved with the KKR-CPA calculations by considering a chemical disorder on sites B, C and D.

\end{abstract}


\maketitle

\section{Introduction}
Sigma ($\sigma$) phase has a tetragonal unit cell (type D$^{14}_{4h}$ P$_4$2/mnm) hosting 30 atoms
distributed over five crystallographically non-equivalent sites.
Due to high coordination numbers (12-15) the $\sigma$-phase is a member of the Frank-Kasper
family of phases.
The importance of this class of structures is reinforced by the fact that they exhibit
topological properties similar to simple metallic glasses due to an icosahedral local
arrangement. Consequently, they can also be regarded as very good approximants for
dodecagonal quasicrystals \cite{Simdyankin00}.
The $\sigma-$phase has been revealed to occur only in such alloy systems in
which at least one element is a transition metal.

Concerning the existence of the sigma-phase in binary alloys, about 50 examples
were found so far \cite{Hall66,Joubert08}.
Its physical properties are characteristic of a given alloy, a high
hardness and brittleness being a common feature. In the family of the binary alloy
sigma-phases only two viz. Fe-Cr and Fe-V have well-evidenced magnetic properties
\cite{Read68,Sumimoto73,Cieslak04,Cieslak08a,Cieslak09}.
In particular, the magnetic properties of $\sigma-$phase strongly depend on composition.
In the Fe$_{1-x}$Cr$_x$ system, where the $\sigma-$phase exist in a narrow range of
$\sim 0.45\ge x \ge \sim 0.50$,
the Curie temperature varies between 0 K (for $x\approx0.50$) and 38 K
(for $x\approx0.45$) \cite{Cieslak04}. In the
Fe-V system, where the range of the $\sigma$-phase occurrence is about six-fold wider, the
effect of the composition is even stronger.
%Its physical properties, in general, depend on the alloy system, a high hardness
%and brittleness being, however, a common feature.

Another common characteristics is a lack of
stoichiometry, a feature that in combination with the complex crystallographic
structure, makes the $\sigma$-phase a very challenging object both for theoretical
investigations and interpretation of experimental data. On the other hand,
the lack of stoichiometry may also have some advantage viz. the $\sigma$-phase
can be formed in a certain range of composition, giving thereby a chance for
studying the influence of content on its properties, hence a possibility of
tailoring these properties. For example, a magnetism of the $\sigma$-phase in
Fe$_{1-x}$V$_x$ alloys is strongly dependent on the composition, and, in
particular, the Curie temperature may vary between 0 K (for $x\approx0.65$)
and 320 K (for $x\approx0.34$) \cite{Cieslak09}. The complex
structure and the chemical disorder (all five sites are occupied by
various atoms) result in a huge number of  possible atomic configurations
which makes an interpretation of measurements performed by using microscopic
methods e.g. Nuclear Magnetic Resonance or M\"ossbauer Spectroscopy to be a
difficult and often non-unique task.
In these circumstances, theoretical calculations, as these presented elsewhere
\cite{Cieslak08b,Cieslak10a}, can be very helpful as they not only provide a deeper insight into the
underlying physics but they also offer a very useful support for a correct
interpretation of experimental data.

This paper reports theoretical calculations on the magnetic properties of the
$\sigma-$FeCr system which is regarded as the archetype of all $\sigma-$ phases.
Its magnetic properties were recently intensively investigated \cite{Cieslak04,Cieslak08a},
and, in particular, the average magnetic moment per Fe atom was determined from magnetization
measurements assuming only Fe atoms having  parallel moments contribute. However, the
knowledge is not complete yet, as it is unknown whether or not both types of atoms
(Fe and Cr) are magnetic, and if so, what are the values of their magnetic moments
on particular sublattices and what is their mutual orientation. These questions cannot
be answered experimentally because one cannot grow big enough single-crystals of the
sigma for a decisive neutron diffraction experiment. Hence, the situation prompts and
justifies further relevant theoretical calculations. The ones reported in this paper are in the
line with this aim.

\begin{figure}[bh]
\includegraphics[width=.49\textwidth]{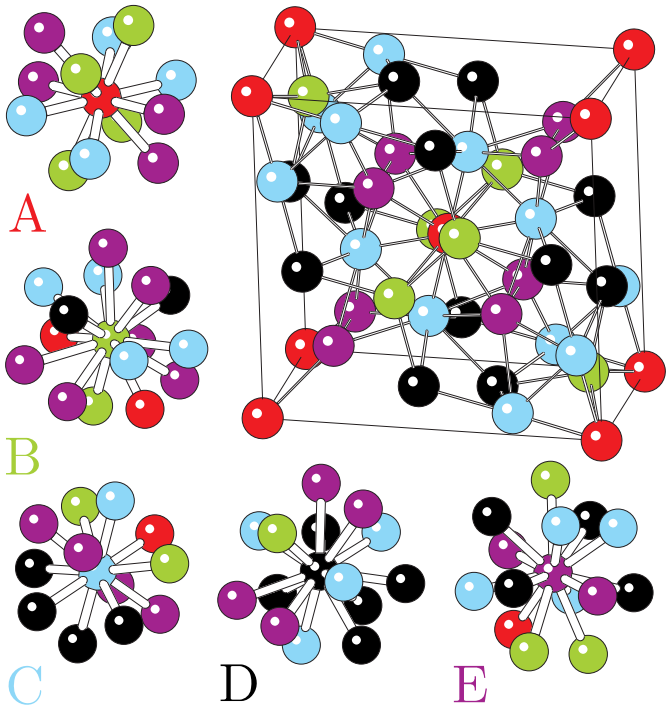}
\caption{(Online color)
Unit cell of the $\sigma$-phase. Atoms belonging to various crystallographic
positions are indicated by
different fillings. NN atoms of each of five lattice sites are indicated also.
}
\label{fig0}
\end{figure}

\section{Computational details}

As aforementioned, the $\sigma$-phase has complex close-packed tetragonal structure with thirty
atoms in the unit cell (Fig.~\ref{fig0}). Atoms are distributed over five non-equivalent sites
in the unit cell, called A, B, C, D and E, the population of them is shown in
Table~\ref{table0}. Each position can be characterized by the total number of nearest
neighbors, NN, their distribution over five sublattices, the distances to NN atoms as well as
the occupancy by Fe or Cr atoms. Some properties can be derived directly from the space
group information. Each atom in the $\sigma$-phase structure has a high coordination number
(from 12 to 15 atoms), belonging to various sites (Table~\ref{table0}).
Their average inter-atomic distances substantially differ and are in the range from
$\sim 2.27~$\AA (the smallest E--E distance) to  $\sim 2.92~$\AA (the largest B--E distance).
Interestingly, the NN spatial distribution is not far from spherical for each site
(Fig.~\ref{fig0}).

Electronic structure calculations for $\sigma-$FeCr compounds have been performed
using the charge and spin self-consistent Korringa-Kohn-Rostoker method. The
crystal potential of the muffin-tin form has been constructed within the local
density approximation (LDA) framework, applying the von Barth-Hedin formula
\cite{Barth72} for the exchange-correlation part. With the self-consistent
crystal potentials converged up to 0.1 mRy and charges up to $10^{-3}~e$,
total, site-decomposed and $l$-decomposed ($s$, $p$ and $d$) density of states
(DOS) have been computed on a 601 energy point mesh for ordered models, employing
the tetrahedron {\bf k}-space integration technique (dividing the irreducible
part of the Brillouin zone into ~120 small tetrahedrons).

In the first stage, the electronic structure calculations have been done for ordered
approximants of the $\sigma-$FeCr alloys, and for that purpose we lowered the symmetry
of the unit cell (space group P$4_2$/mnm) to a simple tetragonal one. In practice,
the tetragonal unit cell and atomic positions were unchanged but variable occupancy
made all 30 atomic positions crystallographically nonequivalent. In such a specified
unit cell each of the crystallographic positions was occupied exclusively either by
Fe or Cr atom. However, in our numerical runs we were constrained by the experimentally
determined Fe/Cr concentrations on each of the five lattice sites \cite{Cieslak08c}
and the composition of the $\sigma-$Fe$_{16}$Cr$_{14}$, being close to the
measured stoichiometry (Fe$_{0.538}$Cr$_{0.462}$), has been considered.

%33333333333333333
It should be noticed that the composition of the alloy is given in two equivalent
notations. When the unit cell of the $\sigma$-phase is referred to calculations,
we use the Fe$_{30-x}$Cr$_x$ formula, since $x$ must be the whole number here.
Conversely, the Fe$_{1-x}$Cr$_x$ formula is related to those cases,
when $x$ (the Cr concentration) in the sample was obtained experimentally.
%33333333333333333

%
\begin{table} % tabela 1
\caption{\label{table0} Atomic crystallographic positions and numbers of the
nearest neighbor atoms, NN, for the five lattice sites of the
$\sigma$-phase.}
\begin{tabular}{|l|l|c|c|c|c|c|c|} \hline
Site& Crystallographic positions & \multicolumn{6}{c|}{ NN}                          \\ \hline
    &           & A   & B   & C   & D   & E  & Total  \\ \hline
A   & 2i (0,     0,     0    )        & -   & 4   & -   & 4   & 4  & 12              \\ \hline
B   & 4f (0.4,   0.4,   0    )        & 2   & 1   & 2   & 4   & 6  & 15              \\ \hline
C   & 8i (0.74,  0.66,  0    )        & -   & 1   & 5   & 4   & 4  & 14              \\ \hline
D   & 8i (0.464, 0.131, 0    )        & 1   & 2   & 4   & 1   & 4  & 12              \\ \hline
E   & 8j (0.183, 0.183, 0.252)        & 1   & 3   & 4   & 4   & 2  & 14              \\ \hline
\end{tabular}
\end{table}

\begin{table} % tabela 1
\caption{\label{table1}
Fe site-occupation parameters of the $\sigma$-FeCr alloy, experimental and
assumed for calculations. $N_t$ stays for the percentage of the total Fe atoms
on the site (referred to the 30 atoms unit cell), whereas $N$ describes the
number of Fe atoms occupying the site. $N_{1}$ and $N_{2}$ represent the values used
in the calculations for ordered (KKR) and disordered (KKR-CPA) models, respectively.
The corresponding relative occupancies (percentage) are given in parenthesis.
}
\begin{tabular}{|l|l|l|l|l|} \hline
Site&$N_t$   &  $N$         &$N_{1}$         &$N_{2}$          \\ \hline
A   & 11.3   &1.826 (91.3)  &2  (100.0)      &2  (100.0)       \\ \hline
B   &  6.4   &1.040 (26.0)  &1  (25.0)       &1  (25.0)        \\ \hline
C   & 20.5   &3.304 (41.3)  &3  (37.5)       &2  (25.0)        \\ \hline
D   & 44.9   &7.208 (90.1)  &7  (87.5)       &8  (100.0)       \\ \hline
E   & 17.0   &2.744 (34.3)  &3  (37.5)       &3  (37.5)        \\ \hline
\end{tabular}
\end{table}

Then, the effects of chemical disorder on the electronic structure have been studied
employing the KKR method with the coherent potential approximation (CPA), which belongs
to well-established techniques used in a description of intermetallic alloys.
In this case, the space group P$4_2$/mnm, defining unit cell of $\sigma-$phase, was
allowed and five nonequivalent sites were occupied as follows:
A ($2i$ site: 100\% Fe),
B ($4f$ site: 25\%Fe and 75\%Cr),
C ($8i$ site: 25\%Fe and 75\%Cr),
D ($8i$ site: 100\%Fe) and
E ($8j$ site: 37.5\%Fe and 62.5\%Cr).
The occupancies of the sublattices intentionally differed slightly from those used for
the ordered approximants (although it is easy to apply precise concentrations in the
KKR-CPA method) due to the fact that we intended to make possible a coherent comparison
between ordered and disordered models. Thus, in the partly disordered model, Fe and Cr
atoms were randomly distributed on three sites, while two other sites were constrained to
be fully ordered. Consequently, this model slightly differs from the experimentally
determined distribution of Fe and Cr atoms in the unit cell of the $\sigma-$FeCr phase.
In real samples a small fraction of chromium atoms was also detected at A and D sites
(see, Table~\ref{table1}). However, we believe that the partly disordered model, forced
by highly time-consuming KKR-CPA calculations, appears as a good approximation of a fully
disordered $\sigma-$FeCr alloy.

In KKR-CPA calculations, the spin-dependent charge and CPA cycles have been performed
self-consistently on the complex energy plane using an elliptic contour (divided into 12
sections with 4 Gaussian quadrature points). The KKR-CPA Green function was
computed on ~75 special {\bf k} point mesh in the irreducible part of the Brillouin zone.
The Fermi energy, $E_{F}$, for all investigated alloys has been determined from the
generalized Lloyd formula \cite{Kaprzyk90}, which permitted us to find $E_{F}$ precisely
from the full derivative of the CPA Green function (without integrating over occupied states).
Like for KKR runs on ordered models of the $\sigma-$FeCr alloy, for finally converged charges
and crystal potentials in the aforementioned disordered model, the KKR-CPA calculations
resulted in total, site-decomposed and $l-$decomposed (with $l_{max}$=2) DOSs computed on
a 251 energy point mesh. More details on the KKR-CPA method can be found in
Refs.\cite{Bansil99, Stopa04}.

In all our calculations, experimental crystallographic data (lattice constants and atomic
positions) were employed \cite{Cieslak08b}.

In order to study the magnetism of the $\sigma-$FeCr alloy spin-polarized KKR as well as
KKR-CPA calculations have been performed. A collinear ordering of magnetic moments
(on Fe and Cr atoms), as allowed by the employed computational codes, were assumed
in a ferromagnetic (FM) state both in the ordered approximants and in the disordered
alloys. However, the FM model of the $\sigma-$FeCr alloy yielded a serious discrepancy
between calculated and experimental values of a total magnetization. Consequently,
a more refined search for achieving a better arrangement of Fe/Cr magnetic moments, with the
use of the symmetry group analysis (see Sec. III C), have been carried out. A relatively
complex model of the magnetic structure, based on anti-parallel alignments of the magnetic
moments within selected sublattices (called APM), was proposed and applied. Finally, spin-polarized
KKR-CPA calculations were also performed for such magnetic unit cell, assuming a collinear
ordering of Fe and Cr magnetic moments, only.

In all figures presented throughout the paper, $E_{F}$  is located at zero, and spin-polarized DOS curves
are given in Ry$^{-1}$ per spin direction.

\section{Results and Discussion}

\subsection{Nonmagnetic state}

\begin{figure}[bt]
\includegraphics[width=.50\textwidth]{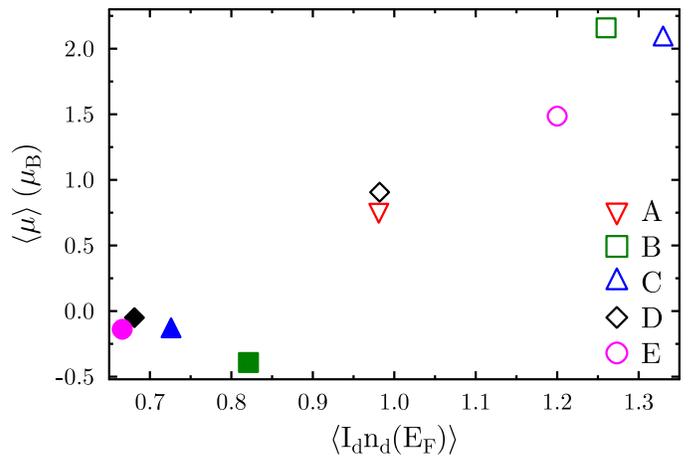}
\caption{(Online color)
Average magnetic moments, $<\mu>$, for five crystallographic sites versus average $<I_d\cdot n_d (E_F)>$-value.
Full symbols stand for Cr atoms, while empty ones for Fe atoms.
}
\label{fig_Stoner}
\end{figure}

We start the analysis of electronic and magnetic properties of the $\sigma-$FeCr from the
spin non-polarized calculations in order to verify the origin of magnetism in view of the Stoner
model. Table ~\ref{table2} presents averaged computed values of $d$-DOS at $E_{F}$, $n_d (E_F)$,
and corresponding Stoner parameter, $I_d$, on all five nonequivalent sites occupied by Fe
and/or Cr atoms. We see that the Stoner criterion ($I_d\cdot n_d (E_F) > 1$)
is satisfied for all iron atoms ($I_d^{Fe}\sim0.036 Ry$), being the highest
one in the case of Fe at B site and the smallest one for Fe at A site. Conversely, the
same condition is not fulfilled for chromium atoms ($I_d^{Cr}\sim0.034 Ry$)
essentially due to much lower DOS at $E_F$, which would suggest that
the magnetic moments appearing on chromium atoms in the $\sigma-$FeCr alloy are in
principle induced due to the presence of rather large values of Fe magnetic
moments within the NN shell. Figure~\ref{fig_Stoner} shows a dependence of the
averaged magnetic moments calculated for the five sublattices versus the
Stoner product that well supports the aforementioned experimental results, suggesting
itinerant character of magnetism in this system \cite{Cieslak08a}.

\subsection{Ferromagnetic state}

\begin{figure*}[bt]
\includegraphics[width=.99\textwidth]{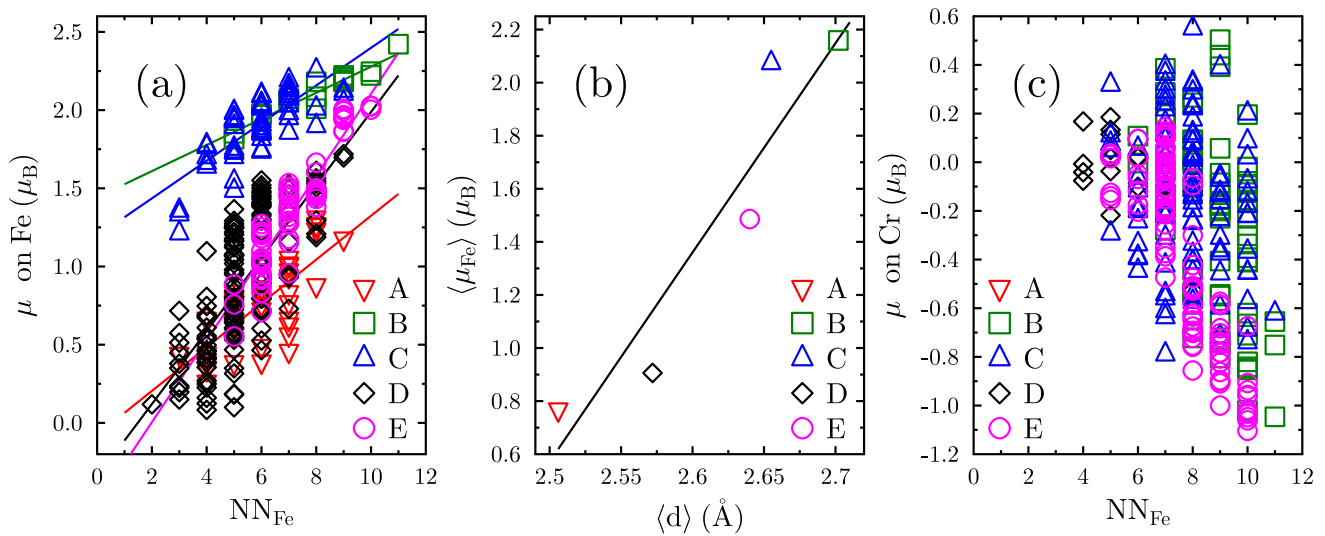}
\caption{(Online color)
Magnetic moments for five crystallographic sites versus the number of $NN_{Fe}$
atoms for (a) Fe and (c) Cr.
The average values of the Fe atom magnetic moments, $<\mu_{I_{Fe}}>$, occupying different sites versus
the average NN-distance, $<d>$, is shown in (b). Solid lines are the linear best-fits to the data.
}
\label{fig1}
\end{figure*}

The aforementioned predictions from the spin non-polarized DOS and the Stoner product analysis
have been followed by spin-polarized KKR calculations in order to determine local and average
magnetic moments on Fe and Cr atoms. The magnetic structure of the
$\sigma-$FeCr has not been determined yet.
Hence, we first discuss the ferromagnetic ordering, being the expected upper
limit of total magnetization.
The values of the magnetic moments of each of 30 atoms in the unit cell obtained
for the ordered approximants are presented in Fig.~\ref{fig1}. One should notice
here a relatively large dispersion of the magnetic moment of Fe atoms,
$\mu_{Fe}$, from almost 0 to nearly $2.5\mu_B$. The values of $\mu_{Fe}$ for each of
the five sublattices are, additionally, nicely correlated with the number of Fe atoms
in the first coordination shell, $NN_{Fe}$. The linear correlation was
described using the equation
\begin{equation}
  \mu_{I_{Fe}}=a_I+b_I NN_{Fe}
\label{eq1}
\end{equation}
where index $I=$A,B,C,D,E denotes the particular sublattice.

The average value of $\mu_{I_{Fe}}$, $<\mu_{I_{Fe}}>$, for each sublattice depends both on
$a_I$ and $b_I$ parameters obtained from Eq.~\ref{eq1} fitted to the data.
Using probabilities of finding particular $NN_{Fe}$-values, $P_I(NN_{Fe})$, determined
from a binomial distribution for each sublattice, $<\mu_{I_{Fe}}>$ can be calculated
with the simple formula
\begin{equation}
 <\mu_{I_{Fe}}> = \sum \mu_{I_{Fe}} P_I(NN_{Fe})
\label{eq2}
\end{equation}
As is clear from Fig.~\ref{fig1}b, $<\mu_{I_{Fe}}>$ almost linearly increases with
the average distances to their nearest atomic neighbors, $<d>$, for all sublattices.
The magnetic moments on Cr atoms are much smaller and mostly oppositely aligned to those of
Fe atoms. Their values range from -1.1 to 0.6 $\mu_B$, and their correlations with the number of
$NN_{Fe}$ are much weaker than those in the case of Fe. For this reason, the average values of
$\mu_{I_{Cr}}$ for various sublattices were obtained as arithmetic means of all
KKR calculated values.

\begin{figure}[h]
\includegraphics[width=.50\textwidth]{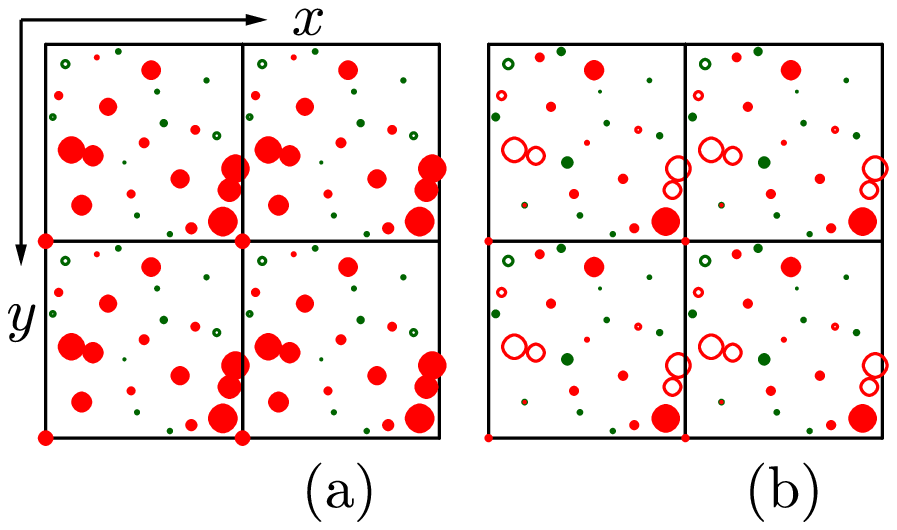}
\caption{(Online color)
Magnetic moments of Fe and Cr atoms in the unit cell, as seen along $z$ axis, for (a) FM structure
and (b) APM (i.e. anti-parallel ordering on D and C sites) structure. Full and empty circles
represent magnetic moments parallel ($\mu_p$) and antiparall ($\mu_a$) to the $z-$direction,
respectively. The size of red (Fe) and green (Cr) circles is proportional to the $\mu$-value.
}
\label{fig2}
\end{figure}

Since, according to the present calculations, both Fe and Cr atoms possess the magnetic moment,
it seems more reasonable to express the
calculated magnetization of the $\sigma$ - FeCr phase, namely the total magnetic moment of the unit
cell, per one atom in the unit cell (i.e. divided by 30) rather than per Fe atom (i.e. divided by 16
in this case).

\begin{figure*}[tb]
\includegraphics[width=.99\textwidth]{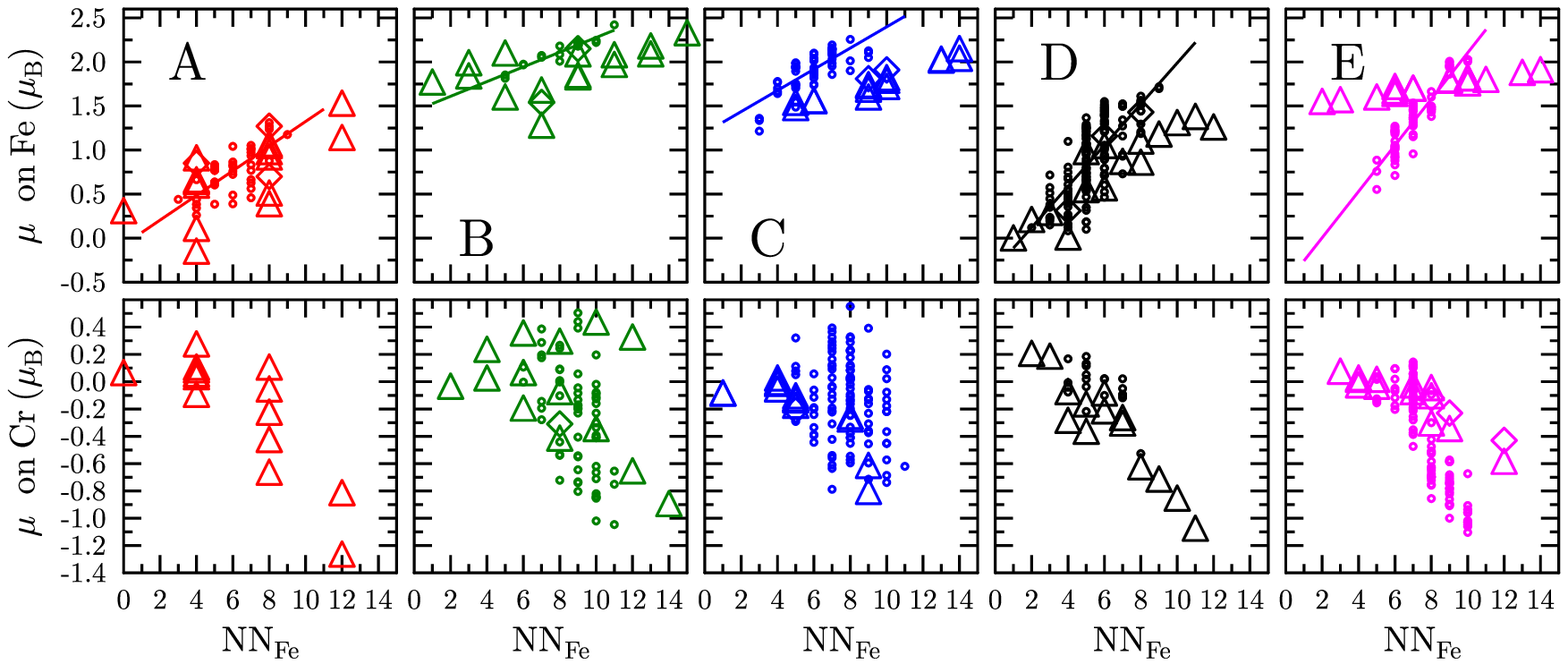}
\caption{(Online color)
Magnetic moments of Fe and Cr atoms for five crystallographic sites versus the number of $NN_{Fe}$
atoms. Circles and solid lines stand for the KKR results (this work), whereas triangles and diamonds
correspond to the calculations reported in Refs. \onlinecite{Pavlu10} and \onlinecite{Kabliman09},
respectively.
}
\label{fig3}
\end{figure*}

The average value of the total magnetic moment of the unit cell can be determined in two ways
(Table~\ref{table2}).

1. As the average value of just all total magnetic moments obtained in each calculation (e.g. 26
different approximants of the $\sigma-$ phase, involving 780 atoms). In this case, we are able to take
into account the spin polarization of electrons located outside the muffin-tin spheres, giving
non-vanishing contribution to total magnetization. This procedure gives $0.55 \mu_B$ for the
magnetic moments inside the muffin-tin spheres and only $-0.02\mu_B$ for the interstitial area
($ 0.53 \mu_B$ in total).

2. As the sum of the weighted five $<\mu_I>$-values, where the weights are the numbers of
atoms belonging to each sublattice (Eq.~\ref{eq2}). This method takes into account all possible
configurations within the NN-shell, even those that have not been considered in the applied model.
Interestingly, the obtained value of $0.59 \mu_B$ is quite close to the previous one.

An illustrative example of a spatial distribution of the magnetic moments of individual atoms in
the unit cell
is shown in Fig.~\ref{fig2}a. In order to present the calculated magnetic structure more clearly,
the $z$-axis projected unit cell was doubled along $x$ and $y$ directions. It is worth noticing that
large magnetic moments (polarized parallel) assemble to some extend. Other analyzed configurations
of atoms (not shown) give qualitatively similar picture.

\begin{table*} % tabela 1
\caption{\label{table2} The average Stoner product, $S_i=<I_d\cdot n_d (E_F)>$, ($i$=Fe,Cr)
in
the $\sigma$-FeCr alloy as calculated for the non-magnetic (NM) state.
Magnetic moments in $\mu_B$ of Fe and Cr atoms as well as their average values, $<\mu>$, are
indicated for various magnetic models and particular lattice sites.
$\mu_{Fe_1}$ and $\mu_{Fe_2} $ in the FM model stand for the values
determined in two ways described in the text. The lack of values for Cr atoms on certain sites
is a consequence of the assumed distribution as presented in Table \ref{table1}.
}
\begin{tabular}{|l|c|c|c|c|c|c|c|c|c|c|c|c|} \hline
Site   &\multicolumn{2}{c|}{NM model} &\multicolumn{4}{c|}{FM model} & \multicolumn{3}{c|}{APM model} & \multicolumn{3}{c|}{APM-CPA model} \\ \hline
       &$S_{Fe}$&$S_{Cr}$&$\mu_{Fe_1}$&$\mu_{Fe_2} $&$\mu_{Cr}$&$<\mu>$ & $\mu_{Fe}$&$\mu_{Cr}$&$<\mu>$ & $\mu_{Fe}$&$\mu_{Cr}$&$<\mu>$    \\ \hline
A      & 0.98 & ---  & 0.77  & 0.76   & ---    & 0.76  & 0.39  & ---    & 0.39  & 0.32       & ---       &  0.32      \\ \hline
B      & 1.26 & 0.82 & 2.16  & 2.13   & -0.27  & 0.33  & 2.07  & -0.11  & 0.43  & 2.08       & -0.10     &  0.44      \\ \hline
C      & 1.33 & 0.73 & 2.08  & 1.91   & -0.13  & 0.63  & 0.07  & -0.06  &-0.01  &-2.06/+2.11 &-0.11/+0.01& -0.07      \\ \hline
D      & 0.98 & 0.68 & 0.91  & 0.90   & -0.04  & 0.78  & 0.24  & -0.01  & 0.20  &-0.44/+0.65 & ---       &  0.21      \\ \hline
E      & 1.20 & 0.67 & 1.49  & 1.38   & -0.39  & 0.28  & 0.99  & -0.20  & 0.25  & 0.90       & -0.25     &  0.18      \\ \hline\hline
av.    &      &      & 1.30  & 1.24   & -0.25  & 0.55  & 0.49  & -0.12  & 0.20  & 0.45       & -0.16     &  0.15      \\ \hline
\end{tabular}
\end{table*}

\begin{figure*}[t]
\includegraphics[width=.99\textwidth]{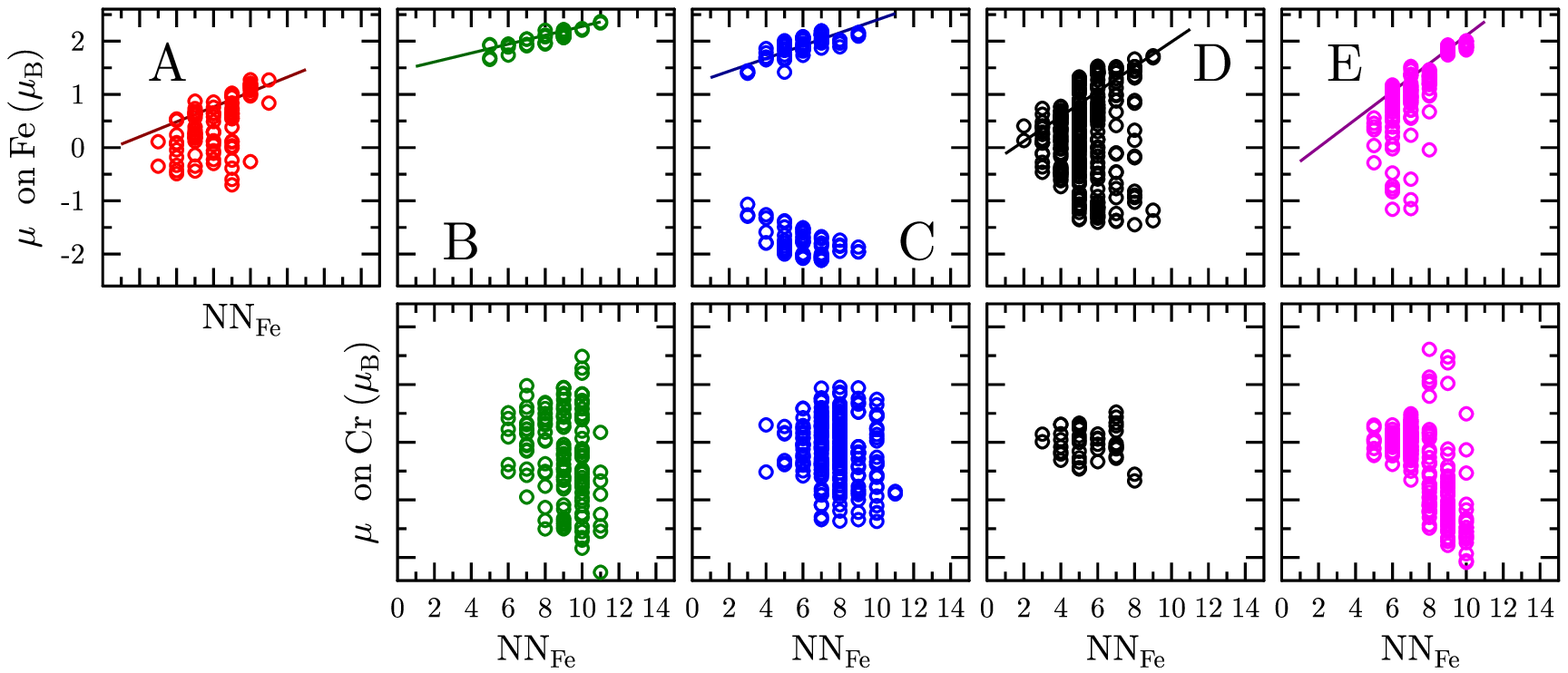}
\caption{(Online color)
Magnetic moments of Fe and Cr atoms for five crystallographic sites versus the number of $NN_{Fe}$
atoms as obtained with the APM model of the magnetic ordering (circles). Solid lines stand
for the best fits to the magnetic moments in the FM ordering shown for the sake of comparison.
}
\label{fig4}
\end{figure*}

Comparable results were obtained in the work of Pavl\.u {\it at al.} \cite{Pavlu10}. The authors performed
spin-polarized calculations using the VASP program, considering only the fully occupied sublattices
by one type of atoms. For a $\sigma-$Fe$_{0.533}$Cr$_{0.467}$ sample the total magnetic
moments (per atom) found by these authors were as large as 0.52, 0.59 and 0.73 $\mu_B$ for three possible atomic
configurations. Applying similar simplifications for atoms distribution in the unit cell,
Kabliman {\it et al.} \cite{Kabliman09} obtained $\mu$=0.75 $\mu_B$ ($\sigma-$Fe$_{0.60}$Cr$_{0.40}$)
and 0.25$\mu_B$ ($\sigma-$Fe$_{0.47}$Cr$_{0.53}$) using FLAPW method.

Due to the fact that these authors \cite{Pavlu10,Kabliman09} reported the average values of the magnetic
moments on each sublattice and also for all considered atomic configurations, their data can be
directly compared with our results.
As can be seen in Fig.~\ref{fig3}, in most cases the KKR values of the magnetic moments are slightly
larger than the corresponding ones reported elsewhere \cite{Pavlu10,Kabliman09}. The observed difference
probably results from a difference in
lattice constants taken for calculations in the present work (experimental data), being slightly
different than those used in Refs. \onlinecite{Pavlu10} and \onlinecite{Kabliman09} (obtained from a relaxation of the unit cell). Indeed,
applying the $a=b$ and $c$ values of the lattice constants from Ref. \onlinecite{Pavlu10}, yielded
smaller magnetic moments. So, the overall agreement with the previous calculations
is satisfying, except for the sublattice E, where the results reported in Ref.
\onlinecite{Pavlu10} remain practically constant ($~1.5~\mu_B$ for Fe), while the corresponding
values obtained in the present paper are clearly $NN_{Fe}-$dependent. Actually, we observe that the magnetic moments
on all five sublattices vary with $NN_{Fe}$ in more or less similar way. However, we are not aware of
any particular reason for markedly different behavior of Fe and Cr magnetic moments on the E site
with respect to the behavior found for other sites, as suggested in
Refs \onlinecite{Pavlu10} and \onlinecite{Kabliman09}.

A good compatibility of the theoretical results obtained for the fully ordered systems of
the $\sigma-$FeCr phase \cite{Pavlu10,Kabliman09} assumed throughout the whole range of concentrations on
the one hand, and those obtained for the ordered approximants of disordered system (corresponding
to realistic configurations for the $\sigma-$Fe$_{0.533}$Cr$_{0.467}$ alloy) on the other, allow us to
deduce the influence of the $NN_{Fe}$ on the hyperfine parameters to be weakly (if at all)
dependent on the alloy stoichiometry.
Therefore, the determined dependences of $\mu_I(NN_{Fe})$ for each sublattice can be assumed
to be valid in the whole range of the $\sigma$-FeCr existence.

The calculated average value of the magnetic moment ($0.59~\mu_B$ per atom per unit cell) should
be referred to the experimental value obtained for the $\sigma-$Fe$_{0.538}$Cr$_{0.462}$,
viz. $\mu=0.14\mu_B$ \cite{Cieslak04}, which is almost four times smaller than the theoretical result.
Such a large and unusual discrepancy between calculated and measured values of the magnetization
on one hand, as well as a non-saturated behavior of $M(H)$-dependence \cite{Cieslak04} on the other,
may indicate the magnetic structure of the $\sigma$-phase is not a simple ferromagnetic one as was
assumed in the aforementioned KKR calculations.

\subsection{Symmetry analysis and models of magnetic ordering}

The large discrepancy between the theoretical (assuming FM model) and experimental results of the total
magnetization in the $\sigma$-FeCr prompted us to search for a possible different magnetic
ordering in such complex phase using the group theory and symmetry analysis as described below.

Let us consider an arbitrary crystal property, $X$, that is defined locally as $\hat{u}$, for all atomic
positions in a crystal with $n$ atoms per unit cell. The local single-site property can be a
scalar, a vector (polar or axial) or even a tensor. According to the symmetry analysis method,
the global quantity $X$ can be written in the most conveniently selected basis, what usually
denotes a symmetry-adapted basis, consisting of the basis vectors  $\Psi_{\nu,\lambda}^{(k_L)}$ of
irreducible representations (IR) $\tau_{\nu,\lambda}^{(k)}$ of the symmetry space group,
with $C_{\nu\lambda}^{(k_L)} $  as linear combination coefficients\cite{Izyumov90},

\begin{equation}
    {X} = \left[
                \begin{array}{c}
                                 %\left[
                                      \begin{array}{c}
                                             \hat{u}(\vec{r_1})  \\
                                             \hat{u}(\vec{r_2})  \\
                                             ...                 \\
                                             \hat{u}(\vec{r_n})
                                      \end{array}                \\
                                 %\right]                                          \\
                    ...                                                           \\
                                 %\left[
                                      \begin{array}{c}
                                            \hat{u}(\vec{r_1}+\vec{t}_{p,q,s})  \\
                                            \hat{u}(\vec{r_2}+\vec{t}_{p,q,s})  \\
                                            ...                                 \\
                                            \hat{u}(\vec{r_n}+\vec{t}_{p,q,s})
                                      \end{array}
                                 %\right]
                \end{array}
          \right]
=\sum_{k_L\nu\lambda}C_{\nu\lambda}^{(k_L)}
                 \left[  \begin{array}{c}
                                 %\left[
                                      \begin{array}{c}
                                            \Psi_{\nu,\lambda}^{(k_L)}(\vec{r_1})  \\
                                            \Psi_{\nu,\lambda}^{(k_L)}(\vec{r_2})  \\
                                            ...                                    \\
                                            \Psi_{\nu,\lambda}^{(k_L)}(\vec{r_n})
                                      \end{array}                                  \\
                                 %\right]                                        \\
                                 ...                                            \\
                                 %\left[
                                      \begin{array}{c}
                                            \Psi_{\nu,\lambda}^{(k_L)}(\vec{r_1})e^{i\vec{k_L}\vec{t}}  \\
                                            \Psi_{\nu,\lambda}^{(k_L)}(\vec{r_2})e^{i\vec{k_L}\vec{t}}  \\
                                            ...                                                         \\
                                            \Psi_{\nu,\lambda}^{(k_L)}(\vec{r_n})e^{i\vec{k_L}\vec{t}}
                                      \end{array}
                                 %\right]
                         \end{array}
                \right]
\end{equation}

\begin{equation}
    \vec{t}_{p,q,s}=p\vec{a}+q\vec{b}+s\vec{c}
\end{equation}
where $l$, $n$ and $\lambda$ correspond to the vectors {\bf k}, to the irreducible
representations of the group and to the IR dimensions, respectively.

In the case of any phase transition, the form of the basis vectors and the information on
the representations relevant to this transition (so-called active representations) are
directly given by the theory of groups and the representations. One is able to calculate these
quantities using a dedicated software\cite{Sikora04}. Then the linear combination coefficients
can be determined by imposing proper conditions on the solution (specific for a given physical
quantity), and the symmetry-adapted ordering modes can be found. The representation $\tau_\nu$
and the coefficients $C_\lambda^{k_l,\nu}$ uniquely determine the symmetry of the structure,
independently of the kind of the property taken into account. The type of the phase transition
and the property under consideration are encoded in the form of the basis vectors.

In this work the symmetry analysis is applied to the structure of the $\sigma$-phase
described by the P4$_2$/mnm space group, where the Wyckoff positions 2$a$, 4$f$, 8$i$, 8$i$' and 8$j$ are
occupied by both Cr and Fe atoms. The aim of our analysis is to find a possible ordering of
the magnetic moments that allow for a magnetic phase transition without any change of
the crystal structure. As the first step, the active representations of the invariable
lattice~{\bf  k}=(0,0,0) for different positions and for different types of ordering (modes)
have been calculated and shown in Table~\ref{table3}. The designations P$_1$, P$_2$ and P$_3$ of
the mode types correspond to a scalar (change of the probability of the site occupations),
polar (displacement of atoms from the equilibrium positions in the high symmetry structure)
and axial (ordering of the magnetic moments) modes, respectively. Hence, for a discussion
of a possible magnetic structure in the $\sigma$-phase, only the latter modes are relevant.
The values presented in Table~\ref{table3} indicate a number of different basis vector sets available
for a given IR. Since the ordering of the magnetic moments is not associated with any change of
the crystal structure in this system, only the models of the magnetic structure belonging to $\tau_1$ IR
are considered.

\begin{table} % tabela 3
\caption{\label{table3} Active representations of all types of modes for P4$_2$/mnm space group
and {\bf k}=0.
The meaning of the symbols is described in the text.
}
\begin{tabular}{|c|c|l|l|l|l|l|l|l|l|l|l|}  \hline
site & mode & \multicolumn{10}{c|}{representation}    \\ \cline{3-12}
%    & type & \multicolumn{8}{c|}{dim 1} & \multicolumn{2}{c|}{dim 2}    \\ \cline{3-12}
     & type &$\tau_1$&$\tau_2$&$\tau_3$&$\tau_4$&$\tau_5$&$\tau_6$&$\tau_7$&$\tau_8$&$\tau_9$&$\tau_{10}$ \\ \hline
     & P$_1$&  1    &  -    &  -    &  -    &  -    &  -    &  1    &  -    &  -    &  -      \\
2$a$ & P$_2$&  -    &  -    &  -    &  1    &  -    &  1    &  -    &  -    &  -    &  2      \\
     & P$_3$&  -    &  -    &  1    &  -    &  1    &  -    &  -    &  -    &  2    &  -      \\ \hline
     & P$_1$&  1    &  -    &  -    &  -    &  -    &  -    &  1    &  -    &  -    &  1      \\
4$f$ & P$_2$&  1    &  -    &  1    &  1    &  1    &  1    &  1    &  -    &  1    &  2      \\
     & P$_3$&  -    &  1    &  1    &  1    &  1    &  1    &  -    &  1    &  2    &  1      \\ \hline
     & P$_1$&  1    &  -    &  1    &  -    &  1    &  -    &  1    &  -    &  -    &  2      \\
8$i$ & P$_2$&  2    &  1    &  2    &  1    &  2    &  1    &  2    &  1    &  2    &  4      \\
     & P$_3$&  1    &  2    &  1    &  2    &  1    &  2    &  1    &  2    &  4    &  2      \\ \hline
     & P$_1$&  1    &  -    &  -    &  1    &  -    &  1    &  1    &  -    &  1    &  1      \\
8$j$ & P$_2$&  2    &  1    &  1    &  2    &  1    &  2    &  2    &  1    &  3    &  3      \\
     & P$_3$&  1    &  2    &  2    &  1    &  2    &  1    &  1    &  2    &  3    &  3      \\ \hline
\end{tabular}
\end{table}

As can be seen from Table~\ref{table3} the ordering of the magnetic moments in this case is possible
only on positions 8i (C and D sites) and 8j (E sites). The results of our calculations are given below:

Site 8i: \\ $\mu_1= (0,0,\mu_0)= \mu_2=\mu_3= \mu_4= -\mu_5= -\mu_6= -\mu_7= -\mu_8   $

Site 8j:
\\ $\mu_1= (\mu_0, -\mu_0,0)= -\mu_2= -\mu_7= \mu_8                   $
\\ $\mu_3= (\mu_0, \mu_0,0)= -\mu_4=\mu_6= -\mu_5                      $

where the numbering of the atomic positions is given according to the International Tables for X-Ray
Crystallography \cite{table_cryst}. The parameter $\mu_0$ is free (both in magnitude and in sign) what means
that the symmetry analysis is not able to predict differences neither in values nor in relative ordering of the magnetic moments
between different sublattices. Due to this freedom different models of possible magnetic orderings
may be considered. However, the calculations clearly show that the arrangements of the magnetic moments
within 8$i$ and 8$j$ sublattices are antiferromagnetic with strictly given sequences, and, additionally,
at 8$j$ site the structure is noncollinear (4 magnetic moments aligned antiparallel in the basic plane along the diagonal and 4 other ones
also antiparallel but perpendicular to the previous moments). It is worth noticing that the purely antiferromagnetic state
suggested by the aforementioned symmetry analysis is possible when atomic positions are occupied by given types of
atoms with the same probability. This type of the lattice occupation is broken in the case of a Fe/Cr disordered alloy, and the proposed magnetic structures
can be treated as initial conditions for the electronic structure calculations. This is the reason why, one of the considered
model of the magnetic structure, has been called APM, instead of antiferromagnetic one.

\subsection{Ordered model of magnetic APM $\sigma$-FeCr}

The KKR calculations using the APM model were performed assuming the input crystal potentials
obtained for the FM state. The computations started with initial spin-polarized potentials,
but this fact had no effect on the finally converged APM results. Such procedure
allowed us to substantially accelerate the self-consistent convergence. According to the APM
model, antiferromagnetic ordering was assumed on the sublattices C and D. The symmetry
analysis does not provide any information about the relative magnetic moments sequences.
These sequences can be known within each sublattice separately. Consequently, two separate
magnetic structures should be considered with the following relative sublattice settings:
C ($\uparrow\downarrow\uparrow\downarrow$) and D ($\uparrow\downarrow\uparrow\downarrow$)
as well as C ($\uparrow\downarrow\uparrow\downarrow$) and
D ($\downarrow\uparrow\downarrow\uparrow$), which led us to different values of the magnetic moments.
Since only collinear arrangement was allowed in the KKR calculations, antiparallel coupling of the magnetic
moments on E site was not taken into account.

As is clear from Fig.~\ref{fig4}, the assumed APM model for the Fe/Cr magnetic moment arrangement
has converged to a stable magnetic configuration with atomic moments of the opposite polarization,
which can be especially well seen in the case of Fe atoms located on sublattices C and D (and also to
some extent on sublattice E). Furthermore, the APM model computations have generated intermediate
Fe magnetic moments with a different alignment, particularly on sublattices A and E, which were not
observed for the FM ordering. There was not revealed any visible change on the sublattice B with respect to FM.
The changes of the magnetic moments of Cr atoms on all sublattices have been found similar to the Fe
ones. The smallest effect has been observed on the sublattice E.

In the case of APM model, the total magnetic moment as determined by averaging the moments of
individual atoms yielded $0.21 \mu_B$, with negligible contribution (c.a. $-0.01 \mu_B$)
from the outside muffin-tin volume, which ultimately gives $0.20\mu_B$ (Table~\ref{table2}).
This value is significantly smaller than that obtained with the FM model ($0.59 \mu_B$), and
much closer to the experimental value ($0.14\mu_B$) \cite{Cieslak04}. This suggests that the
APM-ordering may be responsible for the small value of the magnetic moment deduced from
the magnetization measurements.

A spatial distribution of the magnetic moments in the $\sigma-$phase unit cell found with the APM model
shows a behavior similar to that detected with the FM model (see Fig.~\ref{fig2}b). In this case,
however, the projected areas of the magnetic moments with the same orientation are markedly smaller.

\subsection{Chemically disordered model for the APM ordering}

\begin{figure}[t]
\includegraphics[width=.50\textwidth]{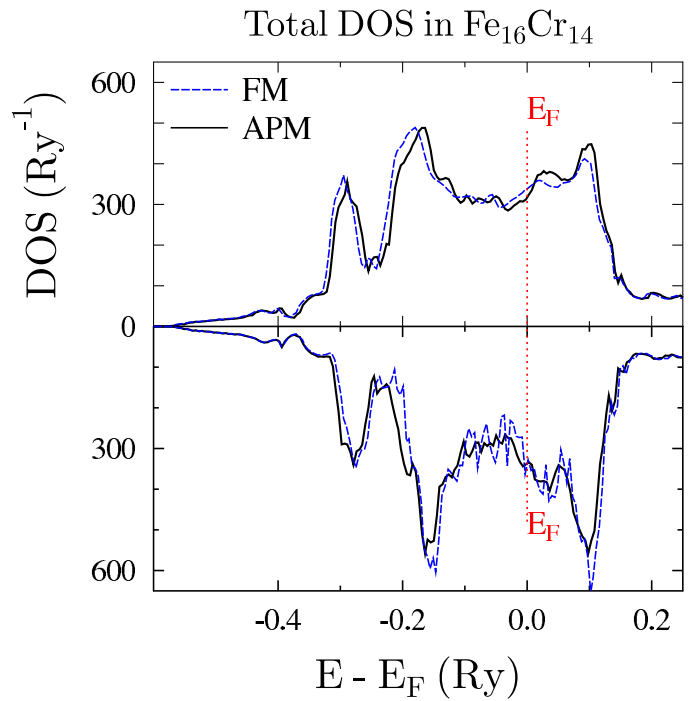}
\caption{(Online color) Total electronic DOS in the $\sigma-$Fe$_{16}$Cr$_{14}$ alloy
obtained with the FM (blue, dashed) and APM (black, solid) models, as calculated by
the KKR-CPA method.}
\label{fig_tdos}
\end{figure}

Bearing in mind the aforementioned KKR results both in the FM and the APM ground
states, we have also studied the effect of a chemical disorder on electronic
structure and magnetic properties of the $\sigma-$Fe$_{16}$Cr$_{14}$ alloy using
the KKR-CPA method.

In this case, the disorder (treated as random) was introduced on three sublattices
(B: Fe$_{0.25}$Cr$_{0.75}$, C: Fe$_{0.25}$Cr$_{0.75}$
and E: Fe$_{0.375}$Cr$_{0.625}$), where the disorder is the most significant
as follows from the neutron diffraction experiment (see Table~\ref{table1}).
Due to the highly time-consuming calculations, the sublattices A and D were
left fully occupied by Fe atoms. This modification required to slightly
decrease the Fe concentration on other sublattices, in order to maintain the
Fe$_{16}$Cr$_{14}$ stoichiometry for the direct comparison between the KKR and
the KKR-CPA results. The KKR-CPA results are discussed here for the FM and the
APM models, neglecting the non-spin polarized computations.

The total DOSs computed for the $\sigma-$Fe$_{16}$Cr$_{14}$ (Fig.~\ref{fig_tdos})
are quite similar in both cases. However, there are few subtle differences
between the two presented DOS spectra. In the FM state, the DOS is more spin polarized
(up-DOS and down-DOS are slightly shifted into lower and higher energy range,
respectively) as well as two down-DOS peaks are more pronounced. This feature
yields the total magnetic moment as large as $\sim0.4 \mu_B$ ($\sim12.0 \mu_B$ per unit
cell) against only $\sim 0.15 \mu_B$ ($\sim4.4 \mu_B$ per unit cell) in the APM
state (Table~\ref{table2}). Interestingly, the DOS at E$_F$ in the APM state is a bit smaller than that
in the FM one, which may tentatively support a better stability of the APM state,
relative to the FM one. Before inspecting in more detail the local origin of
the magnetism in the disordered $\sigma-$Fe$_{16}$Cr$_{14}$ alloy, let us remark
that disorder effects markedly decrease the magnetic moments on Fe atoms on
all sublattices, and, in consequence, the total magnetization per unit cell.
This feature is well seen for the FM state, where the value of 0.4$\mu_B$
was calculated with KKR-CPA for the total magnetization,
compared to almost 0.6$\mu_B$ with KKR as obtained via averaging over a set of results
obtained for the ordered approximant approach. Quite similar behavior was found for the
APM state, since the magnetization values gained from
the KKR-CPA and the KKR methods also differ (0.15$\mu_B$ for disordered alloy against
0.20 $\mu_B$ for ordered models of the $\sigma-$Fe$_{16}$Cr$_{14}$). Hence, an
inclusion of the chemical disorder into the electronic structure calculations
has additionally lowered the total magnetization by about 25\%.
%The effect of disorder on magnetic moments on all sublattices is shown in Table~\ref{ccc}.

\begin{figure}[t]
\includegraphics[width=.50\textwidth]{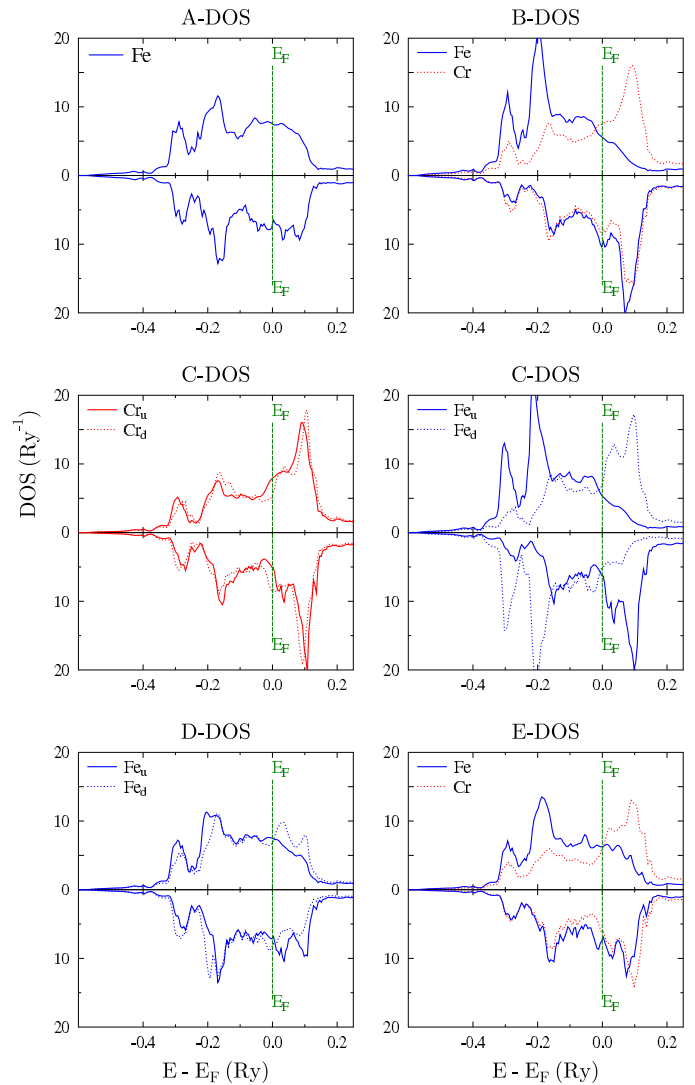}
\caption{(Online color) Site-decomposed KKR-CPA DOS for the $\sigma-$Fe$_{16}$Cr$_{14}$ alloy
as obtained with the APM model. }
\label{fig_dosabc}
\end{figure}

%earing in mind that the choice of the magnetic structure (APM state) appeared

The choice of the APM magnetic structure for the KKR-CPA calculations has resulted in the total
magnetic moment (0.15~$\mu_B$) in an excellent agreement with the experimental data (0.14~$\mu_B$) for the
$\sigma-$Fe$_{0.538}$Cr$_{0.462}$ alloy.
It should be added that in the APM model, according to the symmetry analysis, sites C and D with
multiplicity 8 were split into two "sub-sites", each with multiplicity 4, having
an antiparallel orientation. Consequently, in order to distinguish the magnetic moments and DOSs
of Fe atoms on D site and those of Fe/Cr atoms on C site that are arranged parallel or antiparallel to other sites,
(A, B and E), we have introduced indexes $u$ and $d$, respectively.

Inspecting the Fe DOSs on all sites in the APM state (Fig.~\ref{fig_dosabc}), one can notice two sites
(B and C) with strongly polarized electronic spectrum, giving rise to large
local magnetic moments of +2.08~$\mu_B$ (B site) as well as of -2.06 $\mu_B$ (Fe$_{d}$) and of +2.11 $\mu_B$(
Fe$_{u}$) on C site.
The spin polarization of DOS on other sites remains much smaller, resulting in
relatively lower values of Fe magnetic moments on A (+0.32~$\mu_B$), D (-0.44 on Fe$_{d}$ and +0.65~$\mu_B$
on Fe$_{u}$) and E (+0.90~$\mu_B$). The KKR-CPA results well reflect the aforementioned
correlation (Sec. III B) between the values of the Fe magnetic moment and the average distance to
NN atoms, well supporting the KKR findings for the ordered approximants.
It is worth noticing that on these sites, where Cr atoms were substituted (B, C and E), a small
Cr magnetic moment was found to be always aligned antiparallel to the Fe magnetic moment on
the same site. Such electronic structure behavior was also found for the FM state using the KKR-CPA method.
As already mentioned, the polarization of the Cr DOS is weak and for E site a slight DOS asymmetry
for two spin directions yielded the magnetic moment of about -0.20~$\mu_B$.
In the case of B and C sites, the calculated Cr magnetic moment is as small as -0.10~$\mu_B$, on average.

At present, we must rely on the above described interpretation of the magnetic ordering in the
$\sigma-$FeCr system, since first the APM structure appears the simplest one and second there are
no available experimental evidences indicating more complex (e.g. noncollinear) magnetic structure.

\section{Conclusions}

The results of the electronic structure calculations using KKR and KKR-CPA methods
reported in the present study can be summarized. Both Fe and Cr atoms in the $\sigma-$Fe$_{16}$Cr$_{14}$
alloy are magnetic. Magnetic moments of Fe atoms for each lattice site linearly increase
with the number of Fe atoms in the nearest-neighbor shell, NN$_{Fe}$, whereas Cr atoms are coupled
antiparallel with respect to Fe atoms (values of the magnetic moments for both Fe and Cr atoms
are characteristic of a given lattice site).
Ferromagnetic ordering of the Fe-site magnetic moments accounted for the calculations
results in a significant overestimation of the magnetism in the investigated sample.
Fortunately, admission of an anti-parallel ordering on some selected sites, suggested
by the symmetry analysis, in a combination with a partial chemical disorder yields the
excellent agreement with the experimentally found value of the average magnetic moment
per atom.

\begin{acknowledgments}
This work was partially supported by the Polish Ministry of Science and Higher Education
(MNiSW) under the grant No. N N202 228837.
\end{acknowledgments}

\end{document}